\documentclass[12pt]{article}
\usepackage{graphicx}
\usepackage{epstopdf}
\usepackage{geometry}
\usepackage{mdwlist}
\usepackage{url}
\makeatletter
\g@addto@macro\@verbatim\footnotesize
\makeatother

\title{ PaxosLease: Diskless Paxos for Leases }
\author{ Marton Trencseni, \texttt{mtrencseni@gmail.com} \and
		 Attila Gazso, \texttt{agazso@gmail.com} \and
		 Holger Reinhardt, \texttt{hreinhardt@gmail.com}
}
\date{}

\begin{document}

\maketitle

\abstract{ This paper describes PaxosLease, a distributed algorithm for lease negotiation. PaxosLease is based on Paxos, but does not require disk writes or clock synchrony. PaxosLease is used for master lease negotation in the open-source Keyspace and ScalienDB replicated key-value stores. }

\section{ Introduction }
%%%%%%%%%%%%%%%%%%%%%%%%

In concurrent programming, \emph{locks} are a basic primitive which processes use to synchronize access to a shared resource. In a system where locks are granted without an expiry time (and without a supervisor process), failure of the lock owner before it releases the lock may cause other processes to block.

In a highly-available distributed system, one wishes to avoid single node failures to cause the entire system to block. Also, "restarting" the system in case of failure may be harder than restarting a multi-threaded program. Thus, in distributed systems, \emph{leases} take the place of locks to avoid starvation. \emph{A lease is a lock with an expiry time.} If the owner of the lock fails or becomes disconnected from the rest, its lease automatically expires and other nodes can acquire it.

We assume the basic setup is as follows: the system consists of a set of \emph{proposers} who follow the proposer's algorithm and a set of \emph{acceptors} who follow the acceptor's algorithm. It is assumed the system is non-byzantine, i.e. the nodes do not cheat (are not hacked) by not following their algorithm. The number of acceptors is fixed and does not change.

A naive, majority vote type algorithm can be given which correctly solves the distributed lease problem; \emph{correct} meaning that the lease is held by no more than one node at any time. However, this simple algorithm will frequently \emph{block} in the presence of many proposers, hence the need for a more sophisticated approach. 

The naive majority algorithm is as follows: proposers start local timeouts for $T$ seconds and send requests to the acceptors for the lease with timespan $T$. The acceptors, upon receiving the request, start a timer for $T$ seconds and sends an accept message to the proposer. After the timeout, the acceptors clear their state. If an acceptor receives a request but its state is not empty it either does not answer or sends a reject message. To make sure only one proposer has the lease at any time, the proposer must receive accept messages from a majority of acceptors; then it has the lease until its local timer expires.

As discussed previously, it is possible (and likely), that with many proposers, noone will be able to get a majority and proposers will continually block each other. For example, with three proposers 1, 2 and 3 and three acceptors A, B and C, if the distributed state is such that A accepts 1's request, B 2's and C 3's, then no proposer has a majority. The system must wait until the timeouts expire and the acceptors discard their state at which point the proposers will try again. However, it is likely that the system will block again.

The solution described in this paper is to follow the scheme of Paxos \cite{Parliament}, and introduce \emph{prepare} and \emph{propose} phases, which avoids this kind of blocking altogether\footnote{Another solution may be to let the system block, but introduce an "undo" mechanism which lets some proposers undo their proposals to let one proposer acquire the lease.}. Paxos solves the problem of replicated state-machines, where each node has a local copy of a state-machine, and they wish to reach consensus on the next state transition. Paxos is a majority based algorithm, meaning a majority of the nodes have to be up and communicating for consensus to be possible. Paxos deals with reaching consensus on a single state transition, so in practice several Paxos rounds are run in sequence to negotiate the sequence of state transitions \cite{PaxosMadeLive}. In Paxos, acceptors record their state to disk before sending responses, which guarantees that once a value (state transition) is chosen, it is chosen at all later times; in other words, all state machines go through the same sequence of state transitions, whatever error conditions occur.

Unlike earlier Paxos-based distributed lease algorithms such as Fatlease \cite{Fatlease}, PaxosLease does not make any time synchrony assumptions about local clocks of nodes (no global synchronization is required). Also, Fatlease runs Paxos rounds in succession for lease commands, whereas PaxosLease makes use of the temporal nature of leases and avoids this complication altogether for a simpler and elegant algorithm.

PaxosLease is a natural specialization of Paxos. As in Paxos, it is assumed that the number of nodes is fixed (and node identities are globally known). PaxosLease deals with a special replicated state-machine of the form:

\begin{figure}[htbp]
\begin{center}
\includegraphics[scale=0.8]{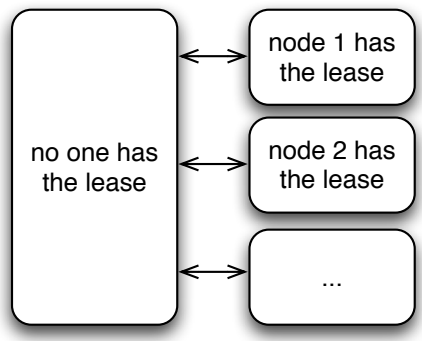}
\caption{PaxosLease distributed state machine.}
\label{default}
\end{center}
\end{figure}

To acquire the lease, PaxosLease proposer nodes propose the value: "node i has the lease", and will automatically return to the "no one has the lease" state after the expiry time. Proposers may also extend their leases by proposing the "node i has the lease" value again before their previous lease expires, and may optionally release the lease before it expires.

Similar to Paxos, PaxosLease intrinsically handles all relevant failure conditions:
\begin{enumerate*}
\item nodes stop and restart
\item network splits
\item message loss and reordering
\item in-transit message delays
\end{enumerate*}

\section{ Definitions }
%%%%%%%%%%%%%%%%%%%%%%%

A PaxosLease cell consists of proposers and acceptors. We assume there are $n$ acceptors and any number of proposers. In practice, nodes often act as proposers and acceptors, but this is an implementation issue and does not affect the discussion.

Proposers send \emph{prepare request} and \emph{propose request} messages to acceptors, who reply with \emph{prepare response} and \emph{propose response} messages. The messages have the following structure:
\begin{enumerate*}
\item prepare request = ballot number
\item prepare response = ballot number, answer, accepted proposal
\item propose request = ballot number, lease
\item propose response = ballot number, answer
\end{enumerate*}

A ballot number and a lease together make up a \emph{proposal}. A \emph{lease} is made up of a \emph{proposer id} (who wishes to become the lease owner) and a timespan $T$.

Acceptors store the following state information:
\begin{enumerate*}
\item highest ballot number promised: acceptors ignore messages whose ballot number is less than this
\item accepted proposal: the last proposal (ballot number and lease) accepted
\end{enumerate*}

There is a globally known maximal lease time $M$. Proposers always acquire leases for timespans $T < M$.
%Proposers guarantee to use different ballot numbers (between each other) and not to-reuse ballot numbers with time interval $M$, the maximal lease time. In practice, this means ballot numbers are divided into two bitfields: one identifies the proposer, the other in a monotonically increasing per-proposer counter which may be reset when the proposer restarts.

Ballot numbers are globally unique and monotonously increasing per proposer. In practice, this can is achieved by ballot numbers being made up of a \emph{proposer id} field, a \emph{restart counter} field and a \emph{run counter} field (at the most significant end) specific to this run of the proposer. The restart counter is incremented each time the proposer starts up and is written to stable storage.

PaxosLease preserves the \emph{lease-invariant}: at any given time, there is no more than one proposer which holds the lease.

\section{ Basic algorithm}
%%%%%%%%%%%%%%%%%%%%%%%%%%

This section describes the basic flow of the algorithm in terms of the proposer's and acceptor's algorithm. The proposers send prepare and propose requests, acceptors reply with prepare and propose responses. If all goes well, it takes two round-trip times for the proposer to acquire the lease.

\begin{enumerate}
\sloppy

\item A proposer wishes to acquire the lease for $T < M$ seconds. It generates a ballot number \texttt{[request.ballotNumber]}, and sends prepare request messages to a majority of acceptors.

\begin{verbatim}
Proposer::Propose()
{
   state.ballotNumber = NextBallotNumber()
   request.type = PrepareRequest
   request.ballotNumber = state.ballotNumber
   Broadcast(request)
}
\end{verbatim}

\item Acceptors, upon receiving a prepare request check whether the ballot number \texttt{[request.ballotNumber]} is higher than their local highest ballot number promised \texttt{[state.highestPromised]}. If it is lower, they either discard the message or send a prepare response with a \emph{reject} answer. If it is equal or higher, the acceptor constructs a prepare response with an \emph{accept} answer containing the currently accepted proposal \texttt{[state.acceptedProposal]}, which may be \emph{empty}. The acceptor sets its highest ballot number promised \texttt{[state.highestPromised]} equal to the ballot number of the message \texttt{[request.ballotNumber]} and sends the prepare response back to the proposer.

\begin{verbatim}
Acceptor::OnPrepareRequest()
{
   if (request.ballotNumber < state.highestPromised)
      return
   state.highestPromised = request.ballotNumber
   response.type = PrepareRespose
   response.ballotNumber = request.ballotNumber
   response.acceptedProposal = state.acceptedProposal  // may be 'empty'
   Send(response)
}
\end{verbatim}

\item The proposer examines the prepare responses from the acceptors. If a majority of acceptors responded with empty proposals, meaning they are open to accept new proposals, the proposer is free to propose itself as the lease owner for time $T$. It starts a timer which will expire in $T$ seconds and sends propose requests containing the ballot number and the lease (its \emph{proposer id} and $T$).

\begin{verbatim}
Proposer::OnPrepareResponse()
{
   if (response.ballotNumber != state.ballotNumber)
      return // some other proposal
   if (response.acceptedProposal == 'empty')
      numOpen++
   if (numOpen < majority)
      return
   state.timeout = T
   SetTimeout(state.timeout)
   request.type = ProposeRequest
   request.ballotNumber = state.ballotNumber
   request.proposal.proposerID = self.proposerID
   request.proposal.timeout = state.timeout
   Broadcast(request)
}

Proposer::OnTimeout()
{
   state.ballotNumber = empty // set in Proposer::Propose()
   state.leaseOwner = false   // set in Proposer::OnProposeResponse()
}
\end{verbatim}

\item Acceptors, upon receiving a propose request check whether the ballot number \texttt{[request.ballotNumber]} is higher than their highest ballot number promised \texttt{[state.highestPromised]}. If it is lower, they can either discard the message or send a propose response with a \emph{reject} answer. If it is equal or higher, the acceptor accepts the proposal: if starts a timeout which will expire in $T$ seconds and sets its accepted proposal to the proposal received (if it has stored a previous proposal, it discards it). It then constructs a propose response with an \emph{accept} answer containing the ballot number \texttt{[request.ballotNumber]}. After the timeout expires the acceptor resets its accepted proposal to \emph{empty}. Acceptors never reset their highest ballot number promised, except when restarting.

\begin{verbatim}
Acceptor::OnProposeRequest()
{
   if (request.ballotNumber < state.highestPromised)
      return
   state.acceptedProposal = request.proposal
   SetTimeout(state.acceptedProposal.timeout)
   response.type = ProposeResponse
   response.ballotNumber = request.ballotNumber
   Send(response)
}

Acceptor::OnTimeout()
{
   state.acceptedProposal = empty
}
\end{verbatim}

\item The proposer examines the propose responses messages. If a majority of acceptors responded accepting the proposal, the proposer has the lease until its local timeout (started in step 3) expires. The time at which the proposer receives the last message necessary for a majority is the time at which it acquired the lease and may switch its internal state to "I have the lease".

\begin{verbatim}
Proposer::OnProposeResponse()
{
   if (response.ballotNumber != state.ballotNumber)
      return // some other proposal
   numAccepted++
   if (numAccepted < majority)
      return
   state.leaseOwner = true // I am the lease owner
}
\end{verbatim}

\end{enumerate}

Note that acceptors do not store their state to stable storage. Upon restart, they start out in a blank state. To make sure restarting nodes do not violate the lease invariant, nodes wait $M$ seconds before rejoining the network. $M$ is a globally known maximal lease time known to all nodes, and proposers always acquire leases for $T < M$ seconds.

\begin{figure}[htbp]
\begin{center}
\includegraphics[scale=0.8]{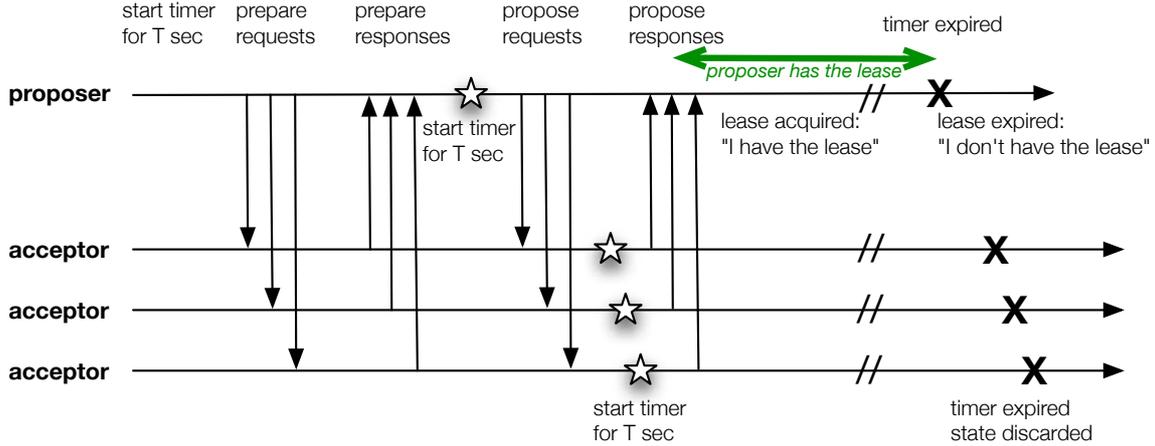}
\caption{Time flow of a proposer acquiring the lease.}
\label{default}
\end{center}
\end{figure}

It is important to note that since all that is transmitted are timespans (relative times), only the proposer which has the lease knows it has the lease. It cannot tell other nodes it has the lease (similar to learn messages in classical Paxos), since other nodes cannot know how much time these learn messages spent in transit. Thus, only the proposer which has the lease knows it has the lease. All the other nodes know is that they do not have the lease. In other words, each proposer has two states regarding the lease: "I do not have the lease and I don't know who has the lease" and "I have the lease". Of course, nodes may send out learn messages as \emph{hints} which may be used in higher-level applications or heuristics, but this is beyond the scope of this paper.

It is possible that a proposer cannot get a majority of acceptors to send favorable responses in steps 3. and 5. above. In this case the proposer may sleep for a while and then restart its algorithm at step 1. with a higher ballot number.

\section{ Proof of lease invariance }
%%%%%%%%%%%%%%%%%%%%%%%%%%%%%%%%%%%%%

First we give the reader an intuitive notion of why PaxosLease works. Figure 2. (above) is a pictorial explanation: the proposer starts its timer before sending out propose requests, acceptors only start their timer \emph{later}, before sending out propose responses, thus, if a majority of acceptors stored the state and started timers, no other proposer will be able to get the lease before the proposer's timer expires. No two proposers will simultaneously believe themselves to be lease holders. 

More formally, PaxosLease guarantees that if proposer $i$ receives propose accept messages from a majority of acceptors for its proposal with ballot number $b$ and timespan $T$ at time $t_{now}$, then if it started its timer at time $t_{start}$, no other proposer will receive a similar majority until $t_{end} = t_{start} + T$.

Proof: Suppose proposer $p$ with ballot number $b$ acquired the lease. It received empty prepare responses of type $accept$ from a majority of acceptors and started its timer at $t_{start}$, received propose responses of type $accept$ from a majority of acceptors at $t_{acquire}$, and thus has the lease until $t_{end} = t_{start} + T$. Let $A_1$ be the majority of acceptors who replied with empty prepare responses to $p$'s prepare requests and $A_2$ be a majority of acceptors who accepted $p$'s proposal and sent prepare responses of type \emph{accept}.

Part 1. No other proposer $q$ can hold the lease between $t_{acquire}$ and $t_{end}$ with ballot number $b' < b$. In order to hold the lease, the proposer $q$ must get a majority $A_2'$ of acceptors to accept its lease. Let $a$ be an acceptor who is in both $A_2'$ and $A_1$: since $b' < b$, $a$ must have first accepted $q$'s proposal and then sent a prepare response to $p$. However, if $a$ sent an empty prepare response to $p$ its state must have been empty, its timer must have expired, thus $q$'s timer also expired, thus $q$ already lost the lease. There is no overlap between $p$ and $q$'s lease.

Part 2. No other proposer $q$ can hold the lease between $t_{acquire}$ and $t_{end}$ with ballot number $b < b'$. In order to hold the lease, the proposer $q$ must get a majority $A_1'$ of acceptors to send it empty prepare responses. Let $a$ be an acceptor who is in both $A_1'$ and $A_2$: since $b < b'$, $a$ must have first accepted $p$'s proposal and then sent a prepare response to $q$. However, since $a$ accepted $p$'s proposal, if it sent an empty prepare response to $q$ its state must have been empty, its timer must have expired, thus $p$'s timer also expired, thus $p$ already lost the lease. There is no overlap between $p$ and $q$'s lease.

\section{ Liveness }

Paxos-type algorithms such as PaxosLease contain the possibility of dynamical deadlock: two proposers can continually generate higher and higher ballot numbers, send prepare requests to acceptors, who continually increase their highest ballot number promised, and neither of the proposers can get acceptors to accept their proposals. In practice, this can be worked around by proposers waiting a small but random time before re-running their algorithm.

The main advantage of Paxos-type algorithms is that static deadlocks, as described in the naive majority vote algorithm, are not possible because proposers can overwrite acceptor states and the algorithm guarantees that no majority is overwritten.

\section{ Extending leases }

In some situations, once a proposer has the lease, it is important that it can hold on to it beyond its original lease time. A typical case is when the lease identifies the master node in a distributed system and it is desirable that a node can hold on to its mastership for a long time.

To accomodate this, only the proposer's algorithm needs to be modified. In step 3., if a majority responded either with empty proposals \emph{or its existing proposal} whose lease has not yet expired, it can propose itself again as the lease holder. This will allow the proposer to extend its lease by $O(T)$ time. The acceptor's algorithm does not change.

\section{ Releasing leases }
%%%%%%%%%%%%%%%%%%%%%%%%%%%%

In the algorithm described so far, the proposers' lease automatically expires after some time. In some situations, it is important to release the lease as soon as possible so that other nodes may acquire it. A typical case is a distributed process, where workers acquire a lease for a resource, do work on it, and then wish to release the lease as soon as possible so that other workers can acquire it.

To accomodate this, any proposer may at any point in time send special release messages to acceptors which contain the ballot number of the lease they wish to release. Before sending out the release messages, the proposer switches its internal state from "I have the lease" to "I do not have the lease". If an acceptor receives a release message, it checks whether its accepted ballot number matches. If if does, it discards its state; otherwise it does nothing. Proposers can also send out release messages to other proposers as hints, telling them that they may be able to acquire the lease now.

\section{ Leases for many resources }
%%%%%%%%%%%%%%%%%%%%%%%%%%%%%%%%%%%%%

The algorithm defines lease actions concerning a single resource $R$. In practice nodes may deal with many resources, such as leases in a distributed process. PaxosLease can be used by running independent instances of the algorithm for each resource by tagging each message, proposer and acceptor state with a \emph{resource identifier}. A node acting as proposer and acceptor requires no more than $\sim 100$ bytes of memory per PaxosLease instance, which means it can handle $\sim 10$ million resource leases per gigabyte of main memory. This and the fact that PaxosLease does not require disk syncs or time synchronization means the algorithm may be used in a wide variety of scenarios for fine-grained locking.

\section{ Implementation}
%%%%%%%%%%%%%%%%%%%%%%%%%

PaxosLease is used for master lease negotiation in the Keyspace replicated key-value store, and its successor ScalienDB. These reference implementations, which contains many practical optimizations, are freely available to all interested readers under the AGPL \cite{AGPL} license at \url{https://github.com/scalien/keyspace} and \url{https://github.com/scalien/scaliendb}.

\section{ Genealogy }
%%%%%%%%%%%%%%%%%%%%%%%%%%%

Leslie Lamport invented Paxos in 1990, but it was only published in 1998. This paper, \emph{The Part-Time Parliament} \cite{Parliament} was "greek to many readers", which led to a second paper, \emph{Paxos Made Simple} \cite{PaxosMadeSimple}. Paxos solves the distributed consensus problem by introducing prepare and propose phases and having acceptors write their state to stable storage before replying to messages. Multiple rounds of Paxos may be run sequentially to negotiate state transitions of a replicated state machine.

Paxos was popularized by its use in Google's in-house distributed stack, described in \emph{Paxos Made Live - An Engineering Perspective} \cite{PaxosMadeLive} and \emph{The Chubby Lock Service for Loosely-Coupled Distributed Systems} \cite{Chubby}. In Google's Chubby, multiple, sequential rounds of Paxos are used to reach consensus on the next write opereration on a replicated database, another way of looking at a replicated state machine.

Fatlease, described in \emph{FaTLease: Scalable Fault-Tolerant Lease Negotiation with Paxos} \cite{Fatlease} solves the same problem as PaxosLease, but it is more complicated as it mimicks the multiple Paxos rounds introduced by the mentioned Google papers, instead of simply expiring the state of acceptors, as in PaxosLease. Also, Fatlease requires the nodes to synchronize their clocks, making it unattractive for real-world use. PaxosLease was inspired by Fatlease and addresses the aforementioned weaknesses.

\end{document}